 \documentclass[final,5p,times,twocolumn]{elsarticle}

\usepackage{hyperref} 
\usepackage{amssymb,amsmath,enumitem}
\usepackage{graphicx}

\usepackage[dvipsnames, usenames]{xcolor}
\definecolor{linkcolor}{rgb}{0.0,0.3,0.5}
\definecolor{darkgreen}{rgb}{0.0, 0.5, 0.0}

\makeatletter
     \renewcommand*\l@figure{\@dottedtocline{1}{1em}{3.2em}}
\makeatother

\usepackage{afterpage}
\usepackage{soul}

\renewcommand{\and}{\\}
\newcommand{\trento}{$\rm T_{R}ENTo$}

\newcommand{\tv}{\trento+v-USPhydro}

\journal{Physics Letters B}

\begin{document}
\begin{frontmatter}

\title{Jet cone radius dependence of $R_{AA}$ and $v_2$ at PbPb 5.02 TeV from JEWEL+$\rm T_RENTo$+v-USPhydro}



\author[first]{Leonardo Barreto}
\author[first]{Fabio M. Canedo}

\author[first]{Marcelo G. Munhoz}
\author[second]{Jorge Noronha}
\author[second]{Jacquelyn Noronha-Hostler}

\affiliation[first]{organization={Instituto de Física, Universidade de São Paulo},
            addressline={C.P. 66318}, 
            city={São Paulo},
            postcode={05315-970}, 
            state={SP},
            country={Brazil}}

\affiliation[second]{organization={Illinois Center for Advanced Studies of the Universe, Department of Physics, University of Illinois at Urbana-Champaign},
            addressline={}, 
            city={Urbana},
            postcode={61801}, 
            state={IL},
            country={USA}}

\begin{abstract}
We combine, for the first time, event-by-event $\rm T_RENTo$ initial conditions with the relativistic viscous hydrodynamic model v-USPhydro and the Monte Carlo event generator JEWEL  to make predictions for the nuclear modification factor 
$R_{AA}$ and jet azimuthal anisotropies $v_n\left\{2\right\}$ in $\sqrt{s_{NN}}=5.02 \, \rm TeV$ PbPb collisions for multiple centralities and values of the jet cone radius $R$. The $R$-dependence of $R_{AA}$ and $v_2\left\{2\right\}$
strongly depends on the presence of recoiling scattering
centers. We find a small jet $v_3\left\{2\right\}$ in mid-central collisions and consistent results in wide jet $p_T$ regions and centralities with ATLAS data.
\end{abstract}


\end{frontmatter}

\section{Introduction} 
Two key signatures of the Quark-Gluon Plasma (QGP), energy loss and collective flow, have been shown to be intertwined in recent years due to the sensitivity of all charged single particle high $p_T$ azimuthal anisotropies, $v_n$, \cite{Wang:2000fq,Gyulassy:2000gk,Shuryak:2001me} to event-by-event fluctuations \cite{Noronha-Hostler:2016eow,Betz:2016ayq,Andres:2019eus}. The smoking gun signature has been a finite value of $v_3$ at $p_T>10$ GeV from \cite{ATLAS:2014qxy,ALICE:2016ccg,CMS:2017xgk,ATLAS:2018ezv,ALICE:2018rtz,ALICE:2018yph} that is only possible through event-by-event fluctuations of the initial conditions \cite{Alver:2010gr,Noronha-Hostler:2016eow,Betz:2016ayq}. However, the impact of these fluctuations on jet measurements remains an emerging and promising field of study, as recent works \cite{Chen:2017zte,Cao:2020wlm,Zhao:2021vmu,He:2015pra,JETSCAPE:2020mzn} have begun to explore the dynamics of jets and event-by-event hydrodynamic backgrounds, improving the description of LHC and RHIC jet data, and new approaches to the jet-medium coupling are now available \cite{Sadofyev:2021ohn,Antiporda:2021hpk}. 

The interplay between jet and hydrodynamic evolution can be further explored by the sensitivity of jet measurements to the jet cone radius $R$ \cite{Tachibana:2017syd,Hulcher:2017cpt,Pablos:2019ngg}, i.e. the anti-$k_T$ jet resolution parameter \cite{Cacciari:2008gp}, since jet distributions are expected to be more affected by the medium response mechanism for increasing $R$ (a clear trend of model-to-data comparisons in \cite{CMS:2021vui,ALICE:2023waz}). Nevertheless, other phenomena, such as color coherence \cite{Casalderrey-Solana:2012evi,Casalderrey-Solana:2011ule,Mehtar-Tani:2011hma,Barata:2021byj} and partonic mass effects \cite{Dokshitzer:2001zm,ALICE:2021aqk,ATLAS:2022agz,Wang:2024yag}, are also crucial to the understanding of the complex $R$-dependence. Therefore, a wide $R$ analysis that includes small and large jets is of direct phenomenological interest for understanding the partonic evolution within the QGP.

Experimental studies have observed the $R$-dependence of various observables, namely $R_{CP}$ \cite{ATLAS:2012tjt}, $\Delta_{recoil}$ \cite{ALICE:2015mdb}, and recently the nuclear modification factor $R_{AA}$ by CMS \cite{CMS:2021vui} and ALICE \cite{ALICE:2023waz}, the latter has applied novel tools to better reconstruct large radii jets with low transverse momentum \cite{Haake:2018hqn,Bossi:2020hwt}. Theoretical calculations have made a number of predictions with simplistic or smoothed hydrodynamic backgrounds \cite{Casalderrey-Solana:2016jvj,Tachibana:2017syd,Chien:2015hda,KunnawalkamElayavalli:2017hxo,Pablos:2019ngg,Mehtar-Tani:2021fud}, mainly the nuclear modification factor and jet spectra. In addition, \cite{JETSCAPE:2022jer,He:2018xjv} studied the $R_{AA}$ of $R \leq 0.5$ jets and \cite{He:2022evt} calculated $v_n$ of small jets using event-by-event hydrodynamic backgrounds. However, theoretical studies that simultaneously investigate the dependence of $R_{AA}$ and $v_n$ with $R$ using state-of-the-art event-by-event fluctuating viscous hydrodynamic backgrounds are still lacking.  

In this work we couple the well-known Monte Carlo parton shower generator JEWEL (Jet Evolution with Energy Loss) \cite{Zapp:2011ya,Zapp:2012ak,Zapp:2013zya} to an event-by-event relativistic viscous hydrodynamic model \tv{} \cite{Moreland:2014oya,Noronha-Hostler:2013gga,Noronha-Hostler:2014dqa,Alba:2017hhe} to investigate $R_{AA}$ and $v_n$ as functions of $R$ for the first time. The more realistic medium approach to the default Bjorken-only expanding smooth medium in JEWEL allows for correlation jet-soft studies, not possible in averaged-out media, and improves significantly jet spectra in non-central collisions. Studying PbPb collisions at $\sqrt{s_{NN}}=5.02 \, \rm TeV$, we find that the $R$-dependence of $R_{AA}$ and $v_2$ strongly depends on the consideration of weakly-coupled medium response as recoiling scattering centers in the simulations. With recoils enabled, $R_{AA}$ increases and $v_2\left\{2\right\}$ decreases with increasing $R$, while the opposite is found without the recoiling option, showing the expected anti-correlation between these observables \cite{Noronha-Hostler:2016eow}. Additionally, we calculate $v_3\left\{2\right\}$ for the first time within the JEWEL framework and a positive triangular flow for $R=0.2$ and $71 < p_T < 398$ GeV jets in mid-central collisions was found, presented with comparisons to ATLAS results \cite{ATLAS:2021ktw}.

\section{Theoretical framework} 
We run a modified JEWEL 2.2.0 \cite{Zapp:2011ya,Zapp:2012ak,Zapp:2013zya} on top of thousands of realistic event-by-event hydrodynamic backgrounds capable of describing soft sector observables \cite{Alba:2017hhe}. JEWEL is a Monte Carlo model that describes the double differential cross-section for the radiation that a parton emits while traversing the medium created in relativistic heavy-ion collisions, thus simulating the parton shower evolution with medium interactions, consistent with the BDMPS-Z formalism \cite{Baier:1996kr,Zakharov:1998sv}. The initial scattering and hadronization processes are performed by PYTHIA 6.4 \cite{Sjostrand:2006za}, while the set of nuclear parton distribution functions EPS09LO \cite{Eskola:2009uj} used is provided by the LHAPDF 5 interface \cite{Whalley:2005nh}. JEWEL was designed specifically to interpolate between the analytically known limits of totally coherent and totally incoherent energy loss regimes, by implementing the destructive interference effect caused by subsequent scatterings in a dense medium \cite{Migdal:1956tc,Zapp:2008af}. The scatterings themselves are treated through a $2\rightarrow2$ elastic cross-section. The radiation is implemented through both initial and final state radiation using the DGLAP formalism of splitting functions. We note that JEWEL is still under active development and the most recent updates \cite{Zapp:2022dhq} are not applied in this study.

JEWEL has successfully described several jet quenching data \cite{Zapp:2012ak,KunnawalkamElayavalli:2017hxo,Andrews:2018jcm,Zapp:2013zya} from LHC in the $2.76 \, \rm TeV$ run. Particularly, it has described the charged hadron $R_{AA}$ quite well from both ALICE and CMS collaborations for 0-5\% collisions. The jet $R_{CP}$ predicted by JEWEL only describes central data and is consistently below the ATLAS results, an effect that becomes more acute in peripheral collisions for $R=0.4$ jets. The $A_J$ distributions agree with CMS data, which indicates a good quantitative description of jet quenching effects. Furthermore, the JEWEL authors suggest that smaller jet cone results are more reliable than larger ones within their framework due to the experimental treatment of the background, which cannot be directly compared to Monte Carlo simulations.

Default JEWEL relies on an ideal Bjorken (0+1D) expanding medium with simplistic transverse profiles given by smooth Glauber initial conditions \cite{Zapp:2013zya}, it models the medium as a collection of scattering centers based only on the \emph{local temperature} and an \emph{ideal gas} equation of state. This simplified model cannot describe spectra and collective flow observables, for which one requires 2+1 or 3+1D event-by-event relativistic viscous hydrodynamics simulations \cite{Niemi:2015voa,Noronha-Hostler:2015uye,Eskola:2017bup,Giacalone:2017dud,JETSCAPE:2020shq,Bernhard:2019bmu,Nijs:2020roc,Denicol:2018wdp,Schafer:2021csj}. Hence, even though out-of-the-box JEWEL allows for event-by-event simulations of the jet evolution, it does not incorporate event-by-event fluctuations of the medium as well, which limits its ability to simultaneously describe soft and hard heavy-ion observables. In this work, we improve this situation by coupling JEWEL to a modern and realistic approach to the medium description. 

In order to couple JEWEL to an arbitrary hydrodynamic event,  the following modifications were implemented \cite{Canedo:2020xzf, Barreto:2021fbt}: 
\begin{itemize}[leftmargin=*]
\item \emph{Initial vertex choice:} The original algorithm implemented in JEWEL limited the hard scattering vertex to be picked from the calculation of the Glauber overlap region between colliding nuclei, considering a probability proportional to the density of nucleon-nucleon binary collisions $n_{coll}$ \cite{Zapp:2013zya}, which is not compatible with an external medium. Instead, we sampled the vertex using $n_{coll}$ in the transverse plane given the medium profile at the start of the hydrodynamic evolution $\tau_0$. By applying a realistic lattice QCD based equation of state and \trento{} entropy deposition mechanism (both as used in the hydro calculations), a map between initial temperature and $n_{coll}$, thus vertex position probability, is created. 

\item \emph{Local fluid velocity:} The description of the scattering centers' momenta was changed to consider the local transverse velocity, provided by the v-USPhydro calculations, instead of JEWEL original thermal-only one. Moreover, the rate of interactions between the evolving parton and scattering centers, modeled as an effective medium density, had to be updated with the factor $\frac{p_\mu u^\mu}{p_0}$, where $p^\mu$ is the parton 4-momentum and $u^\mu$ is the medium 4-velocity \cite{Baier:2006pt, Liu:2006he}. In agreement with \cite{Armesto:2004vz}, which showed how flow from hydrodynamic backgrounds can affect jet observables, the presented implementation resulted in, approximately, a 7\% (most peripheral) to 30\% (most central) increase of $v_2\{2\}$ and a less intense  $R$-dependence of $R_{AA}$ when compared to previous simulations with the same media without local flow effects.
\end{itemize}

JEWEL implements the possibility of recoiling scattering centers as a weakly-coupled approach to medium response \cite{KunnawalkamElayavalli:2017hxo}. This option, referred to as \emph{with recoils}, enables the scattering center to be removed from the medium due to the interaction with a parton, changing its 4-momentum. The recoiled parton free-streams until hadronization and its initial 4-momentum is subtracted directly from the jet constituents following the \emph{Constituents Subtraction} prescription \cite{Milhano:2022kzx}, since thermal contributions are expected to be removed as background in experimental analyses. We present the results of simulations with and without the recoil methodology, as they both were successful in describing different observables \cite{KunnawalkamElayavalli:2017hxo, Milhano:2022kzx}, although inconsistent conclusions regarding experimental data predictability were observed \cite{ALICE:2023jye, ALICE:2023qve}, and display divergent $R$-dependence behavior \cite{CMS:2021vui, ALICE:2023waz, ALICE:2022vsz}. 

For simulations of the medium, we use the standard \trento{} initial condition parameters  $p=0$, $k=1.6$, and $\sigma=0.51 \, \mathrm{fm}$ found using a Bayesian analysis \cite{Bernhard:2016tnd}. Hydrodynamics begins at $\tau = 0.6$ fm/$c$ where only a constant $\eta/s\sim 0.05$ is used. The lattice QCD based equation of state  PDG16+/2+1[WB] from \cite{Alba:2017mqu} is used. This framework  successfully fits (and has made successful predictions for) data from RHIC and the LHC \cite{Alba:2017mqu,Giacalone:2017dud,Rao:2019vgy,Sievert:2019zjr} but ultracentral collisions are still slightly off \cite{Carzon:2020xwp} (this remains as an unresolved issue in the field \cite{Giannini:2022lbj}). 

\section{Experimental observables}
Given the differential jet yield $N_{jet}$ in $N_{evt}$ heavy-ions (AA) collisions and differential jet cross section $\sigma_{jet}$ for pp interactions, the nuclear modification factor \cite{Connors:2017ptx} is defined as
\begin{equation}
  R_{AA}(p_T) = \frac{1}{\langle T_{AA} \rangle} \frac{\frac{1}{N_{evt}} \frac{d^2 N_{jet}}{dp_T dy}\big|_{AA}}{ \frac{d^2 \sigma_{jet}}{dp_T dy}\big|_{pp}}, \nonumber
\end{equation}
where the ratio is scaled by the nuclear thickness function $\langle T_{AA} \rangle = \langle N_{coll} \rangle / \sigma_{NN}$, with $\sigma_{NN}$ as the total inelastic nucleon-nucleon cross section, and yields are normalized considering the simulated sampling cross section \cite{Andronic:2020ygi}. Within the JEWEL framework, the average number of binary nucleon collisions $\langle N_{coll} \rangle$ is set to one for all centrality classes, as it only simulates one dijet production per event \cite{Zapp:2013vla}. Moreover, pp results were calculated evolving JEWEL parton shower in vacuum.

Expanding on JEWEL original tuning methodology \cite{Zapp:2012ak, KunnawalkamElayavalli:2016ttl}, both the scaling factor of the Debye mass $s_D$ $^1$ and the decoupling temperature of the QGP $T_C$ were chosen minimizing the $\chi^2$ of anti-$k_T$ $R = 0.4$ jet $R_{AA}$ data in central 0-10\% PbPb collisions compared to ATLAS results \cite{ATLAS:2018gwx} for each of the models with or without recoils. In contrast to JEWEL default $s_D=0.9$ and $T_C=170 \rm \, MeV$ \cite{Zapp:2013vla}, the best values obtained to fit the experiment using realistic medium profiles were $s_D=1.1$ and $T_C=170 \rm \, MeV$ for both recoil options. The tuning parameters are kept constant throughout all simulations, including peripheral systems and different (pseudo-)rapidity cuts. We treat each of the models independently in the tuning process to eliminate any bias in their predictability of experimental measurements, as the addition of recoiled partons changes the underlying physics of the calculated jet spectra and, consequently, its description of the tuning data. This differs from the original JEWEL tuning, in which the parameters were set without considering recoils.

For anisotropic flow coefficients $v_n$, we assemble histograms describing the azimuthal distribution for each $p_T$-centrality bin. Following the calculations for high-$p_T$ \cite{Noronha-Hostler:2016eow, Betz:2016ayq} and heavy-flavor \cite{Prado:2016xbq} hadrons, the $R_{AA}(p_T, \phi)$ distribution can be calculated as
\begin{equation}
  \frac{R_{AA}(p_T, \phi)}{R_{AA}(p_T)} = 1 + 2\sum_{n=1}^{\infty}v_n^{jet}(p_T) \cos(n(\phi - \Psi_n^{jet}(p_T))),\nonumber
\end{equation}
where $\Psi_n^{jet}(p_T)$ is the reconstructed jet symmetry plane angle
\begin{equation}\label{eq:psin}
  \Psi_n^{jet}(p_T) = \frac{1}{n} \tan^{-1} \left( \frac{\int_0^{2 \pi} d \phi \sin(n\phi) R_{AA}(p_T, \phi)}{\int_0^{2 \pi} d \phi \cos(n\phi) R_{AA}(p_T, \phi)} \right),\nonumber
\end{equation}
and
\begin{equation}\label{eq:vn}
  v_n^{jet}(p_T) = \frac{1}{2 \pi} \int_0^{2 \pi} d \phi \cos(n(\phi - \Psi_n^{jet}(p_T))) \frac{R_{AA}(p_T, \phi)}{R_{AA}(p_T)}.\nonumber
\end{equation}
As pointed out in \cite{Luzum:2013yya}, experimentally, only expected values of $\langle v_n^A v_n^B\rangle$ can be measured (where $A$ and $B$ denote different objects or kinematics). Therefore, each jet $\phi-\psi_n$ in the histogram is weighted by the soft $v_n$ coming from the hydro calculations and the experimental results should be compared to \cite{Noronha-Hostler:2016eow, Betz:2016ayq} 
\begin{equation}\label{eq:vncorr}
  v_n\{2\}(p_T) = \frac{\langle v_n^{soft}v_n^{jet}(p_T) \cos(n(\Psi_n^{soft} - \Psi_n^{jet}(p_T))) \rangle}{\sqrt{\Big \langle \left(v_n^{soft} \right)^2 \Big \rangle}},\nonumber
\end{equation}
for the \emph{jet-soft correlation}, which follows the same form of the 2-particle correlation, with $\langle ... \rangle \doteq \frac{\sum_i M_i R_{AA}(p_T)_i (...)}{\sum_i M_i R_{AA}(p_T)_i}$ being the average over each hydrodynamic event $i$ with multiplicity $M_i$. We emphasize that this observable requires information about soft sector quantities which the JEWEL original treatment of the medium cannot not provide and, thus, the modifications presented in this letter are essential for performing event-by-event studies of jet anisotropies in this model. 

The $R_{AA}$ was calculated oversampling 1000 \tv{} medium profiles per centrality class with 500 simulated hard scatterings each. Differently, 100 media per centrality with, approximately, $1.5 \cdot 10^5$ hard scatterings each were used for the simulations of $v_n$, due to the convergence of $\Psi_n^{jet}(p_T)$ demanding highly populated histograms.
Only the statistical uncertainty of the observables were considered and the jackknife resampling technique was used to estimate the errors of $v_n$ and its ratios (see \cite{Andronic:2020ygi, Zapp:2012ak} for discussions on uncertainties in the model). 

The kinematic cuts and centrality classes were chosen to better suit the comparison to ATLAS results for each observable. The nuclear modification factor was calculated for jets with $|y_{jet}| < 2.8$ and transverse momenta between $50$ and $630 \, \rm GeV$ for 0-10\%, 10-20\%, 20-30\%, 30-40\%, 40-50\%, and 50-60\% centralities, following \cite{ATLAS:2018gwx}. The elliptic and triangular anisotropic flow coefficients, expanding on \cite{ATLAS:2021ktw}, were analyzed for $|y_{jet}| < 1.2$ and $p_T$ in the range of $71$ to $650 \, \rm GeV$, with integrated $p_T$ bins of $71$ to $398 \, \rm GeV$ and $200$ to $650 \, \rm GeV$, for 0-5\%, 5-10\%, 10-20\%, 20-40\%, and 40-60\% centralities. Additionally, ratios of $R_{AA}$ were computed for $|\eta_{jet}| < 2.0$ and $400 < p_T < 500 \, \rm GeV$, centralities of 0-10\%, 10-30\% and 30-50\%, to compare with CMS $R$-varying measurements \cite{CMS:2021vui}. The anti-$k_T$ jet algorithm was applied with $R=0.2, 0.3, 0.4, 0.6, 0.8, 1.0$ for all observables without changes in the (pseudo-)rapidity cuts.
We used FastJet \cite{Cacciari:2011ma} for constructing the anti-$k_T$ jets and Rivet \cite{Bierlich:2019rhm} for developing the presented analyses.

\begin{figure}
  \centering
  \includegraphics[keepaspectratio, width=\linewidth]{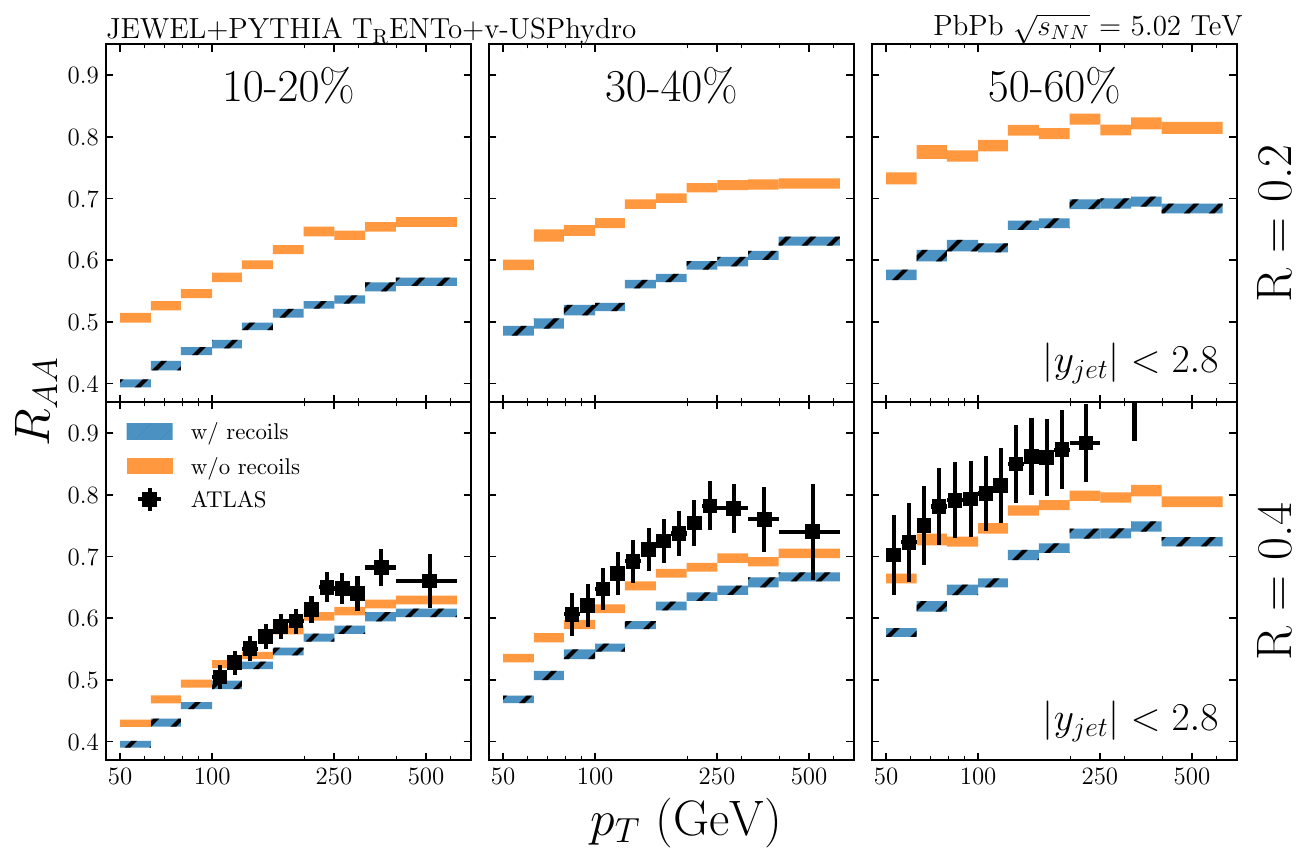}
  \caption{Jet nuclear modification factor for jets with $R=0.2$ (top) and $0.4$ (bottom) compared to ATLAS results \cite{ATLAS:2018gwx} for multiple centralities from left to right: 10-20\%, 30-40\% and 50-60\%.
  }
  \label{fig:raa_ATLAS}
\end{figure}

\section{Results}

With the significant upgrades we made to JEWEL to run event-by-event hydrodynamic backgrounds, our first consistency check is the calculation of $R_{AA}$. In Fig.\ \ref{fig:raa_ATLAS} our model predictions are shown compared to ATLAS $R = 0.4$ data at $5.02 \, \rm TeV$, complemented with $R = 0.2$ results. We find that for central collisions of 10-20\%, \tv+JEWEL works well compared to experimental data. For more peripheral collisions (30-40\% and 50-60\%), the model predicts significantly more suppression than is measured experimentally when recoils are included, but can reasonably reproduce $R_{AA}$ without recoils.

\begin{figure}
  \centering
  \includegraphics[keepaspectratio, width=\linewidth]{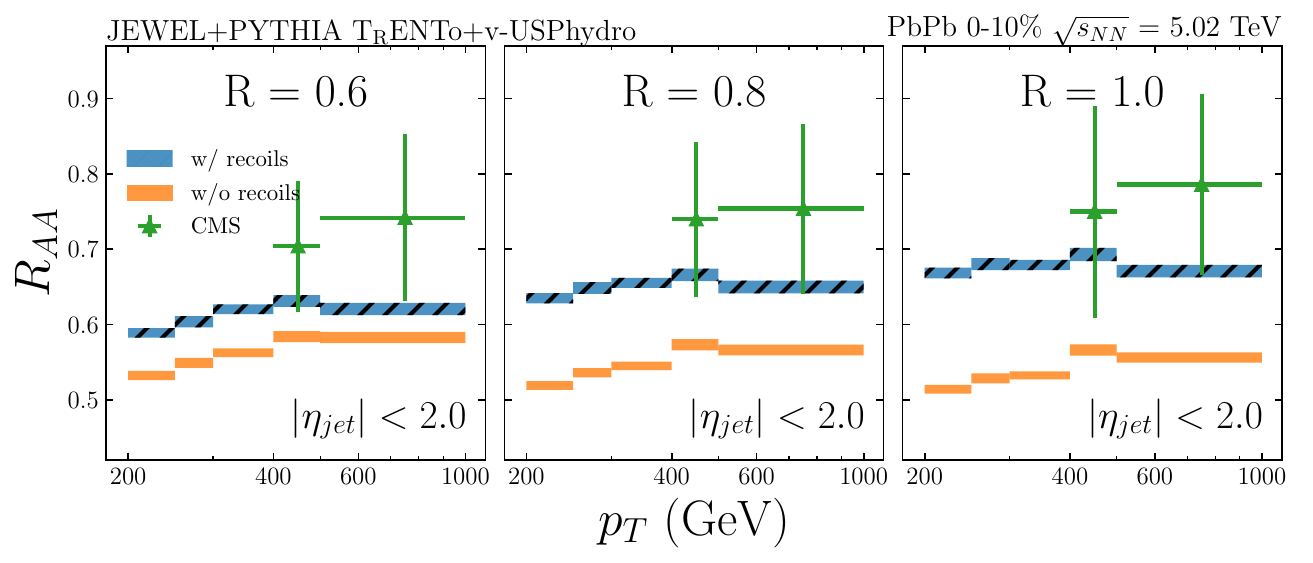}
  \caption{Jet nuclear modification factor for large jets in 0-10\% centrality compared to CMS results \cite{CMS:2021vui} for multiple radii from left to right: 0.6, 0.8 and 1.0.
  }
  \label{fig:raa_large}
\end{figure}

Calculations of the model for large-radius jets $R = 0.6$, $0.8$ and $1.0$ in central 0-10\% collisions against CMS data are presented in Fig. \ref{fig:raa_large}. Unlike small jets, the inclusion of recoils is necessary for the description of the experimental results and implies in a slight increase in $R_{AA}$ with the jet resolution parameter whereas the opposite is observed when medium response is not considered. Although the qualitative trend is consistent with the unmodified JEWEL results in \cite{CMS:2021vui}, the hydrodynamic approach yields in a milder $R$-dependence of large jet $R_{AA}$.

Moreover, the calculations for $R = 0.2$ and $0.4$ show a stronger suppression when recoils are considered. This effect contrasts the expectations of the parton shower model, as the jets should recover part of energy lost with interactions with the medium by the addition of recoiling scattering centers \cite{He:2015pra, KunnawalkamElayavalli:2017hxo, JETSCAPE:2022jer}, thus resulting in a higher $R_{AA}$. The inversion is not observed for calculations with partonic jets with subtracted thermal momenta, in which the hadronization process is skipped, for any jet $R$ and collision centrality presented in this letter for both \tv{} and the default Bjorken-only expanding medium. The effect is caused by the addition of recoils, i.e. medium partons, in the \emph{Lund string formation} as part of the usual hadronization mechanism within JEWEL \cite{KunnawalkamElayavalli:2017hxo} \emph{after the shower evolution} and implies a non-monotonic relation between shower parton-medium interactions and jet suppression that is more prevalent in smaller jets, as seen in the first row of Fig. \ref{fig:raa_ATLAS}, which shall be explored in further work. Large $R \geq 0.6$ jets exhibit no inversion in Fig. \ref{fig:raa_large} as the expected increase in $R_{AA}$ resulting from medium response dominates, due to larger jet area enclosing more of the energy recovered by recoils.
    
We emphasize that the phenomenon is not exclusive to the hydrodynamic modifications and can be replicated in the default medium approach\footnote{For JEWEL 2.4.0 using the usual initial temperature parameter $T_i = 590 \, \rm MeV$, the inversion occurs around the 10-20\%, 40-50\% and 70-80\% centralities classes for, respectively, $R = 0.2$, $0.3$ and $0.4$. The effect becomes more noticeable the more peripheral the system is or lower $T_i$.} but it is \emph{accentuated} by media with characteristics that result in lower average number of interactions, such as \tv{} profiles, lower initial temperatures, or peripheral collisions.

\begin{figure}
  \centering
  \includegraphics[keepaspectratio, width=\linewidth]{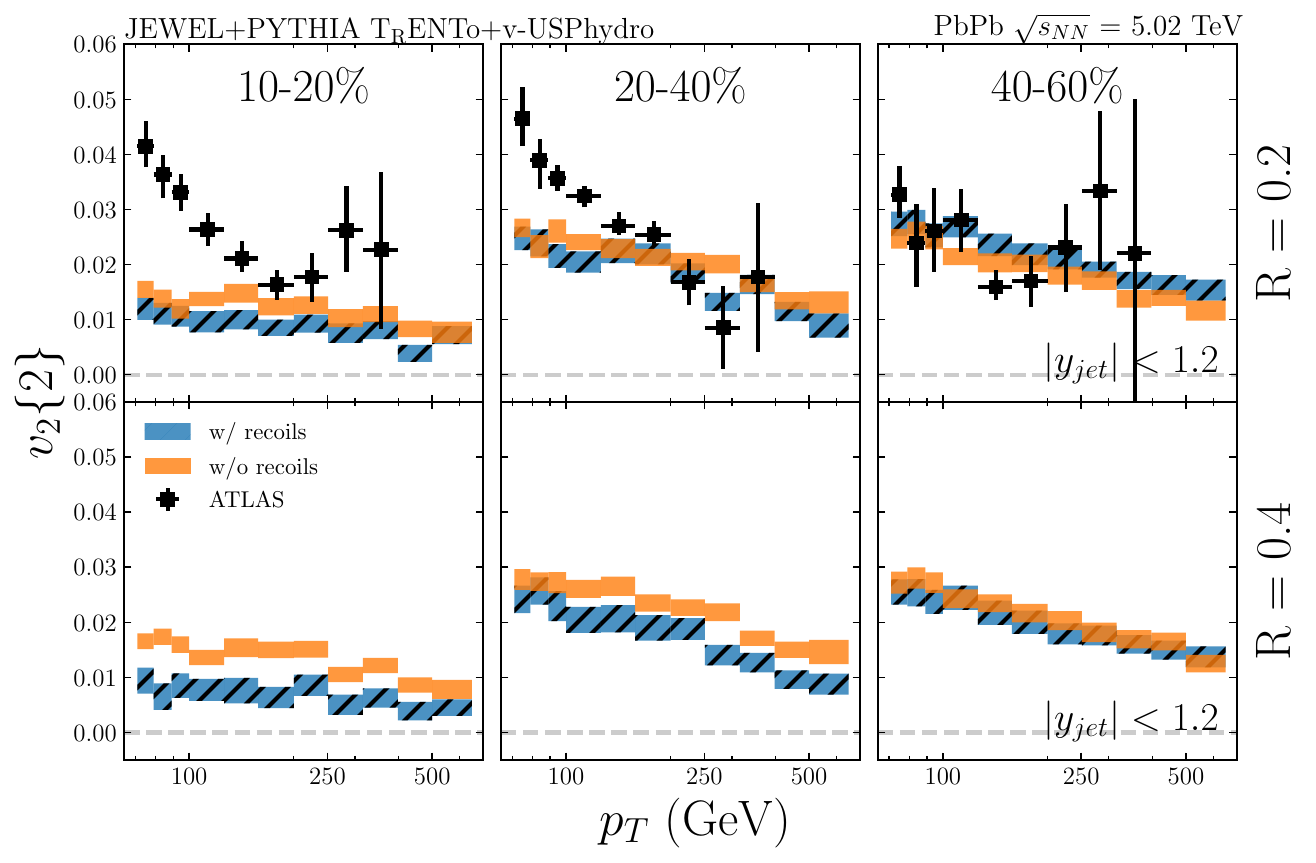}
  \caption{Elliptic flow $v_2\{2\}$ for jets with $R=0.2$ (top) and $0.4$ (bottom) compared to ATLAS results \cite{ATLAS:2021ktw} for multiple centralities from left to right: 10-20\%, 20-40\% and 40-60\%.
  }
  \label{fig:vn_ATLAS}
\end{figure}

In Fig.\ \ref{fig:vn_ATLAS}, we compare calculations of $v_2\left\{2\right\}(p_T)$ in 10-20\%, 20-40\%, 40-60\% centralities with ATLAS $R = 0.2$ data, in addition to $R = 0.4$ results. Our simulations manage to reproduce the ATLAS measurements reasonably well for the most peripheral centrality across all $p_T$. Central and mid-central collisions are lower than data below $p_T \lesssim 150$ GeV but can reasonably reproduce $v_2\left\{2\right\}$ for $p_T \gtrsim 150 $ GeV. Both models display only a minor change when comparing the $R=0.2$ and $0.4$ curves.
    
Regarding the impact of medium response, the jet-soft correlated elliptic flow displays a mild dependence on the description of recoils with increasing $R$, which is aligned with the expectation for small jets but contrasts with the $R_{AA}$ in Fig. \ref{fig:raa_ATLAS}. The addition of recoils implies in a lower or equivalent $v_2\left\{2\right\}$, except in the 40-60\% centrality, that was not observed in \cite{He:2022evt} and may be caused by the finer details of JEWEL recoil implementation, such as the no re-interaction of recoiling scattering centers after their removal from the medium or the lack of hydrodynamic propagation of their holes.

\begin{figure}
    \centering
    \includegraphics[keepaspectratio, width=\linewidth]{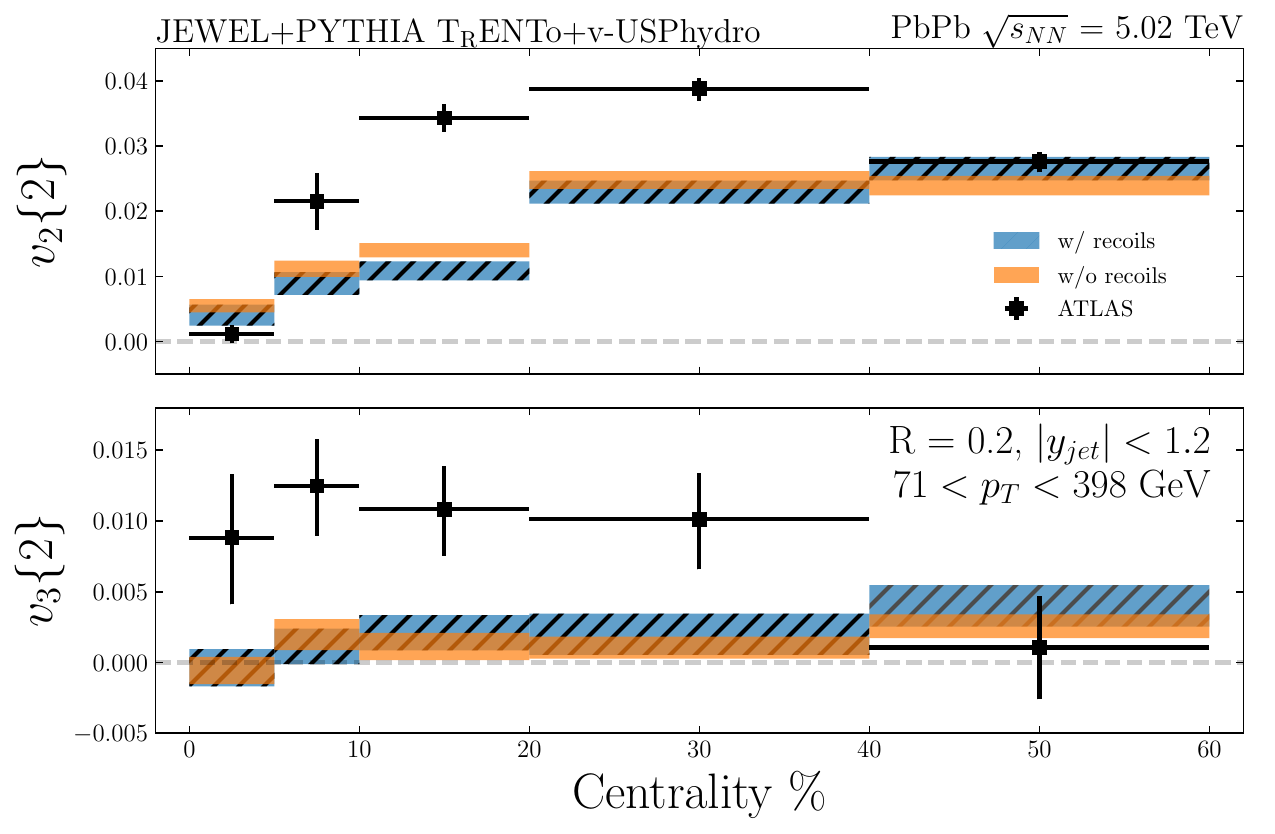}
    \caption{Integrated $v_n\left\{2\right\}$ with $n=2,3$ for jets with $R=0.2$ compared to ATLAS results \cite{ATLAS:2021ktw}. Kinematics cuts of $71<p_T<398$ GeV and $|y_{jet}|<1.2$ are used.
    }
    \label{fig:v2v3cent}
\end{figure}

In Fig.\ \ref{fig:v2v3cent}, we now consider the integrated values of both $v_2\left\{2\right\}$ and $v_3\left\{2\right\}$ across centrality compared to the ATLAS data. In our simulations, we find that the \emph{event-by-event fluctuations} lead to a finite, positive, but small $v_3\left\{2\right\} (\sim 0.003$) except for the most central data, in which the result is consistent with zero. The integrated $v_2\left\{2\right\}$ does well compared to the most central and the most peripheral experimental results, but fails to describe the experimental centrality dependence. For mid-central we obtain slightly lower values for both $v_2\left\{2\right\}$ and $v_3\left\{2\right\}$. Again, we find that these observables generally do not depend strongly on the recoil assumptions, as expected of $R = 0.2$ jets.

\begin{figure*}
    \centering
    \includegraphics[keepaspectratio, width=\linewidth]{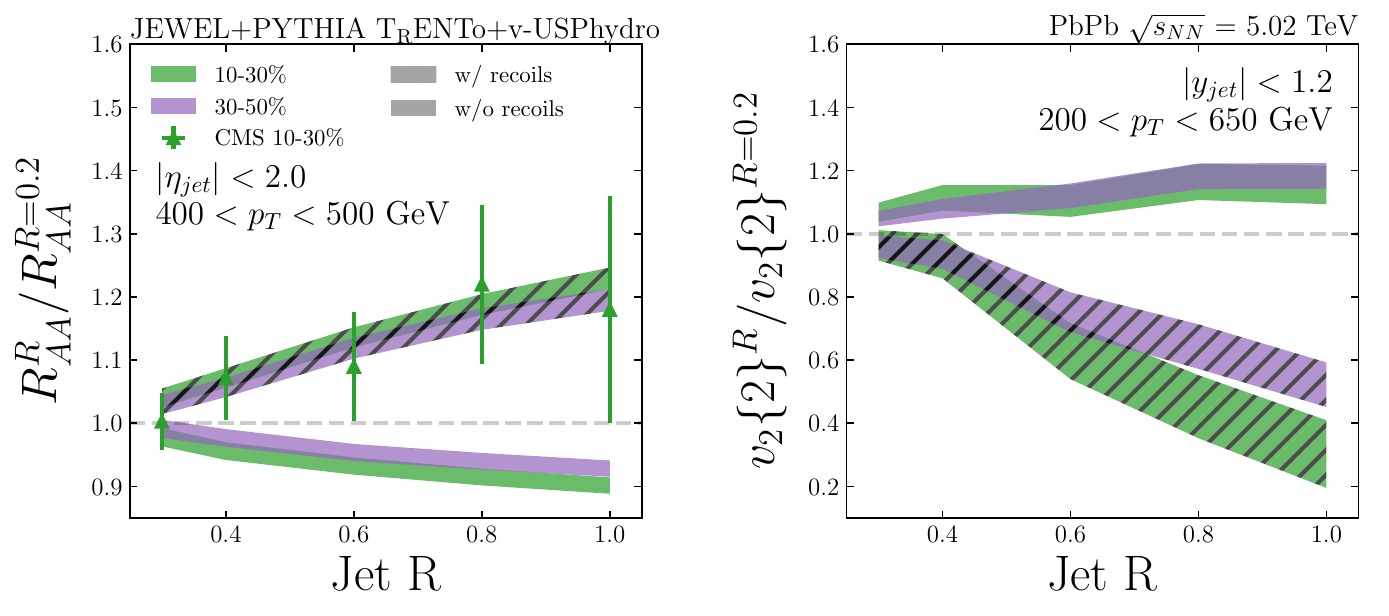}
    \caption{Integrated $R_{AA}$ (left), compared to CMS data \cite{CMS:2021vui}, and $v_2\{2\}$ (right) dependence on the jet cone radius for 10-30\% and 30-50\% centralities. Kinematic cuts of $400<p_T<500 \, \rm GeV$ and $|\eta_{jet}|<2.0$ (left), and $200<p_T<650 \, \rm GeV$ and $|y_{jet}|<1.2$ (right).}
    \label{fig:ratio}
\end{figure*}

Finally, we study the dependence of $R_{AA}$ and $v_2$ with $R$, in the context of the so-called $R_{AA}$ to $v_2$ puzzle\ \cite{Liao:2008dk,Betz:2014cza,Xu:2014tda}. Thus, we use the centralities $10-30\%$ and $30-50\%$ and plot ratios of $R_{AA}$ and $v_2\left\{2\right\}$ vs $R$ in Fig.\ \ref{fig:ratio}. Note that for $R_{AA}$ we use the $p_T$ cuts consistent with CMS from \cite{CMS:2021vui} whereas, for $v_2\left\{2\right\}$, we use the cuts consistent with ATLAS \cite{ATLAS:2021ktw}. Our primary motivation for this is to ease comparisons with potential experimental data (other kinematic cuts are available upon request). With recoils, regardless of the centrality class, we find that $R_{AA}$ increases with increasing $R$ while $v_2\left\{2\right\}$ decreases with increasing $R$. In contrast, the exact opposite effect for our simulations without recoils is observed. The behavior is more acute in central collisions, as jets are more modified by the medium. Currently the only existing data for full jets comes from CMS \cite{CMS:2021vui}, which appears to be consistent with our results \emph{with recoils}\footnote{Recent ALICE results \cite{ALICE:2023waz} for charged jets in a lower $p_T$ interval indicate a decreasing trend in the $R_{AA}$ ratio, but the presented calculations are not directly comparable to them.}.

We note that the results from Figs.\ \ref{fig:raa_large} and \ref{fig:ratio} appear to be in tension with our previous comparisons to data in Figs.\ \ref{fig:raa_ATLAS}, \ref{fig:vn_ATLAS}, and \ref{fig:v2v3cent}. For $R=0.2,0.3 \text{ and } 0.4$ values, i.e. jets with small cones, the option without recoils provides the best fit to experimental data across all centralities and $p_T$ while $R \geq 0.6$ results' indicate a preference for the description with recoils for $p_T \geq 400$ GeV in central collisions. However, for the $R$-dependence, there is a clear preference for the results with recoils. That being said, the $R$-dependent quantities are normalized by the $R=0.2$ values, which implies that while simulations with recoils cannot correctly capture the $R=0.4$ measurements as consistently as without recoils, the current recoiling methodology can correctly determine the $R$-dependence of the jet. Thus, it may be that the medium response mechanism is important for jets with wide radii but is not needed for describing the small-jets' distributions, in agreement with \cite{KunnawalkamElayavalli:2017hxo, He:2018xjv, He:2022evt, Tachibana:2017syd, Pablos:2019ngg}.

We have checked the $R$-dependence of both default JEWEL (see \cite{CMS:2021vui}) and the realistic medium of \tv{} only incorporating the temperature dependence of the profiles \cite{Barreto:2021fbt}. The general $R$-dependence has the same qualitative behavior as Fig.\ \ref{fig:ratio} but default JEWEL largely misses the quantitative dependence with $R$ that our full simulations from \tv{}, including temperature and flow dependence alongside the improved initial vertex selection, are able to capture. Thus, our results indicate that the simultaneous investigation of the $R$-dependence of $R_{AA}$ and $v_2\left\{2\right\}$ provides nontrivial information that can shed light on the complex multiscale processes underlying jet-medium interactions.

\section{Conclusions}
We have performed significant upgrades to the JEWEL jet event generator that allowed for the first calculations of JEWEL coupled to realistic event-by-event relativistic viscous hydrodynamic backgrounds. This new framework is used to make predictions for the dependence of $R_{AA}$ and $v_n\left\{2\right\}$ with the jet cone radius. The model describes the $R_{AA}$, $v_2\left\{2\right\}$, and $v_3\left\{2\right\}$ experimental data across $p_T$ and centralities from ATLAS  reasonably well, especially without using the recoil option. We find that integrated flow harmonics are slightly below the data in mid-central collisions and the results with recoils for the $R_{AA}$ are too low in non-central collisions. At first glance, this would appear to imply that including realistic media decreases the need for recoils in JEWEL. However, we note that the $R$-dependence measurements of $R_{AA}$ from CMS are in tension with that statement, since we find that recoils enhances $R_{AA}$ with increasing $R$ that is consistent with CMS data. The model-to-experiment comparison demonstrate the need of the introduced modifications and \tv{} \emph{realistic hydrodynamic} profiles to describe the observable.
In contrast, simulations without recoils decrease $R_{AA}$ with increasing $R$. 

Our results for the $R$-dependence of the observables are always normalized by the corresponding $R=0.2$ result such that it appears that a wide jet $R \ge 0.6$ requires the physics of recoils but smaller $R < 0.4$ does not. One potential solution to this tension may be that medium response is required for large-area jets' calculations of observables. The $v_2\left\{2\right\}$ has the opposite behavior as $R_{AA}$ when scaling with $R$, regardless of the inclusion of recoils or not. Our study highlights the importance of further $R$ measurements (including $v_2\left\{2\right\}$ measurements at large $R$) to understand if recoils are required and also to see if experiments find the opposite  behavior with $R$ displayed by $R_{AA}$ and $v_2\left\{2\right\}$.

The v-USPhydro interface code for JEWEL and presented Rivet analyses are publicly available at, respectively, \href{https://github.com/leo-barreto/USP-JEWEL}{\texttt{github.com/leo-barreto/USP-JEWEL}} and \href{https://github.com/leo-barreto/USPJWL-rivetanalyses}{\texttt{github.com/leo-barreto/USPJWL-rivetanalyses}}.

\section{Acknowledgements}
The authors would like to thank V.~Bailey, A.~Sickles, I.~Kolbé, and R.~K.~Elayavalli for discussions. M.G.M., L.B. and F.M.C. were  supported by grant \#2020/04867-2, São Paulo Research Foundation (FAPESP). M.G.M. and F.M.C. acknowledge the support from Conselho Nacional de Desenvolvimento Científico e Tecnológico (CNPq) as well. J.N. is partially supported by the U.S.~Department of Energy, Office of Science, Office for Nuclear Physics under Award No. DE-SC0021301. J.N.H. acknowledges the support from the US-DOE Nuclear Science Grant No. DE-SC0020633 and the support from the Illinois Campus Cluster, a computing resource that is operated by the Illinois Campus Cluster Program (ICCP) in conjunction with the National Center for Supercomputing Applications (NCSA), and which is supported by funds from the University of Illinois at Urbana-Champaign. This study was financed in part by the Coordenação de Aperfeiçoamento de Pessoal de Nível Superior – Brasil (CAPES) – Finance Code 001.

\bibliography{bibliography.bib}
\markboth{Bibliography}{}
\bibliographystyle{elsarticle-num}

\end{document}